\documentclass[aps,showpacs,showkeys,superscriptaddress]{revtex4}

\usepackage{amsmath}

\usepackage{amsthm,amssymb}
 \usepackage{makeidx,epsfig,lscape}
 \usepackage{color,colortbl}
 \usepackage{xcolor,pict2e}
\usepackage{amsfonts}
\usepackage[all]{xy}
\usepackage{verbatim}
\def\be{\begin{equation}}
\def\ee{\end{equation}}
\def\bea{\begin{eqnarray}}
\def\eea{\end{eqnarray}}

\definecolor{gray}{rgb}{0.5,0.5,0.5}
 \def\frac#1#2{{#1\over #2}}

 \def\frac#1#2{{#1\over #2}}

\def\be{\begin{equation}}
\def\ee{\end{equation}}
\def\ba{\begin{eqnarray}}
\def\ea{\end{eqnarray}}

\usepackage{graphicx}

\usepackage{comment}

\begin{document}
\title{Holographic Renormalization with Machine learning}

\author{Eric Howard}\email{eric.howard@mq.edu.au}
\affiliation{Macquarie University, Department of Physics and Astronomy, Sydney, Australia}
\affiliation{Griffith University, Centre for Quantum Dynamics, Brisbane, Australia}
\date{\today}

\pacs{71.55.-i; 72.80.Vp; 73.22.-f; 73.63.-b}

\keywords{Renormalization Group, AdS/CFT correspondence, holographic principle, Restricted Boltzmann machines, deep learning}

\begin{abstract}
At low energies, the microscopic characteristics and changes of physical systems as viewed at different distance scales are described by universal scale invariant properties investigated by the Renormalization Group (RG) apparatus, an efficient tool used to deal with scaling problems in effective field theories. We employ an information-theoretic approach in a deep learning setup by introducing an artificial neural network algorithm to map and identify new physical degrees of freedom. Using deep learning methods mapped to a genuine field theory, we develop a mechanism capable to identify relevant degrees of freedom and induce scale invariance without prior knowledge about a system. We show that deep learning algorithms that use an RG-like scheme to learn relevant features from data could help to understand the nature of the holographic entanglement entropy and the holographic principle in context of the AdS/CFT correspondence.
\end{abstract}

\maketitle

\section{Introduction}

Deep learning is a large set of methods using multiple layers of representation to automatically learn relevant features from structured data. Machine learning derives (`learns') patterns from data in order to make sense of previously unknown inputs. At the border between artificial intelligence, neuroscience and statistics, deep learning methods such as deep neural networks, deep belief networks \cite{ackley, zhang, osindero} and recurrent neural networks use learning data representations to process, predict, optimize and discover new features and patterns in big data. Machine learning techniques become useful when unknown features yet to be discovered are far too complex for standard numerical modelling methods. Recent progress in the field of machine learning \cite{tensorflow, lee, neal} has shown significant promise in applying ML tools like classification or pattern recognition to identify phases of matter \cite{tanaka1, wetzel1, whu} or non-linear approximation of arbitrary functions using neural networks. The algorithms can be applied to classical statistical physics problems in two dimensions and generate constraints that are further interpreted as sufficient regularisation conditions for a critical system. Our paper provides a connection between the properties of many-body quantum states and neural network based machine leaning.
Restricted Boltzmann machines (RBM)\cite{norouzi, hinton100, hinton200, mni, sim, tie, aoki, Hinton5, Hinton7} attracted great interest from researchers in the field of quantum many-body physics. The algorithms are based on artificial neural networks used to discover interesting patterns in the input data. 
The neural net is a trainable quantifiable functional mapping with multiple variables. RBM represents a good example of the strong connection between statistical physics and machine learning.
Carleo and Troyer \cite{CT} trained an RBM like wavefunction to study quantum many-body systems away from equilibrium. Deng \cite{DLS} developed an exact RBM model for multiple topological states. Huang and Wang \cite{HM, LW} developed a recommender system RBM-based in order to boost Monte Carlo modelling of quantum many-body systems. Torlai and Melko \cite{melko10} used an RBM for modelling the thermodynamics of statistical physics models. 
Quantum many-body problems assume an exponential scaling of the Hilbert space dimension with the system size therefore requiring a large amount of information computationally daunting and intractable. 
The construction of efficient representations for physical states that access only partially the Hilbert space makes possible the use of much less numerical resources.

The Renormalization Group (RG) plays a key role in our understanding of both quantum field theory and statistical physics. RG is a key tool of condensed-matter theory and particle physics, containing several techniques: real-space RG, functional RG, density matrix renormalization group (DMRG) and many others. A low-energy effective theory with universal characteristics can be deducted from finding and identifying de degrees of freedom that are physically relevant to the theory. Since de discovery of critical systems \cite{Morningstar}, asymptotic freedom in quantum chromodynamics and Kosterlitz-Thouless phase transition, the Renormalization Group methods model the effective behaviour of systems depending on the scale at which they are observed. 

Most systems that are dissimilar at microscopic level display surprisingly similar properties at long distances when observed at a macroscopic scale. The success of RG in studying critical processes is a consequence of the presence of
a smaller number of expanding directions (one or two) at a fixed point of an infinite-dimensional dynamical system, while all other directions become contracted. RG framework exploits these fixed points and the associated universal features in order to identify the long characteristics of diverse physical systems. Unfortunately, most thse systems have several symmetries which makes the deep earning task more computationally demanding. Deep learning promises an unprecedented opportunity in solving complex quantum many-body problems. The advantage of deep multi-layer architectures over shallow architectures and their great success in predicting features in structured data is still not understood.

In this work, we find similarities between the conceptual framework of Renormalization Group \cite{beny1, schwab1, venka} on one hand and deep learning on the other, where depth and scale play similar roles. RG predicts the behaviour of a system as dependent on a scale or energy parameters while here, the same parameter is treated as depth. The deep learning algorithm adapts a simulation problem to a learning task, generating a representation of a classical probability distribution which optimizes the likelihood of the training data, with no prior knowledge of RG or quantum states. The RG concept deals with the transition between different scales and the scale-invariant emergent characteristics of a system.

The motivation of this work is the universal invariant behaviour that is displayed at different scales by diverse systems in nature related to dynamics, symmetries or renormalization, human brain or society. The brain displays power law patterns at both architectural and functional levels. Deep learning has been very successful and has helped progress the state-of-the-art  in several areas ranging from computer vision and pattern recognition and speech recognition to drug  discovery, genomics and computer games. The deep mechanisms that generate the emergent pattern structure and the underlying scale invariance in complex systems involve universal aspects and characteristics like self similarity, criticality or power law distribution. 

The predictive success of the scaling theory supports the idea that close to criticality the correlation length represents the only important length scale, while the microscopic lengths become irrelevant. The critical behaviour will be generated by the fluctuations that are statistically self-similar up to that scale.

We employ a procedure known as the Renormalisation Group in momentum space, developed by Wilson \cite{Wilson} to study systems displaying scale invariant behaviour. We find a one-to-one connection between RBM-based Deep Neural Networks and a genuine field theory by mapping a fully connected graph to a field theory described by a Hamiltonian density. RG in momentum space traces a path in the coupling space for the Hamiltonian and its couplings. The differential equations of the couplings become constraints for the connection weights of the system. 

The assumption of self-similar patterns will eventually eliminate any correlated degrees of freedom at specific length scales up to the point where the system contains simple uncorrelated degrees of freedom at a renormalised length scale.

Deep neural networks can describe the most complex correlations and they are a fine tool to map quantum many-body states. We here study the equivalence between the Renormalisation Group and deep learning. The quantum many-body problem with large degrees of freedom has become a significant and challenging predicament in condensed matter theory. The quantum many-body states may be translated into an efficient computational form using neural networks. By mapping a deep neural network (DNN) to the Renormalization group and identify a sequence of RG transformations to the AdS spacetime, the holographic principle may be better understood. Deep neural networks are neuroscience-inspired statistical models described by multiple layers of neurons, with each layer receiving inputs from the layer below. As a neural network extracts features from complex data, its mapping to the Renormalization group can help with identify features in complex quantum many-body problems. 

Holographic principle was conjectured to solve the black hole information paradox and states that a quantum theory with gravity must be describable by a boundary theory. The principle is a consequence of black hole thermodynamics, as a result of the fact that the entropy of a black hole is proportional to the area of the event horizon, in Planck units. 

The information in the bulk is completely stored on the spacetime boundary. We find that the emergence of a holographic approach on quantum gravity is connected to deep learning processes associated with the partition function of a quantum field theory. We will find out that the mechanism of the holography can b understood in terms of deep learning.
According to the holographic principle all information in the bulk of the spacetime is encoded at its boundary. In this sense the AdS/CFT correspondence \cite{maldacena1, hooft} is a realization of the holographic principle. The partition function in AdS gives us the correlation functions of the boundary CFT. AdS/CFT correspondence is employed to compute the two point functions at the boundary and the RG transformations of the CFT are identified with the AdS geometry.
Our goal is to provide an alternative way to understand the entanglement structure of the spacetime \cite{vidal, swingle, beny} as encoded in the graph of a deep neural network and show that deep neural networks can explain the holographic nature of the spacetime.
We already know that the entanglement structure of the spacetime has a Ryu-Takayanagi \cite{RT} form in AdS/CFT but we find that the entanglement structure of a deep neural network also has a Ryu-Takayanagi form. Spacetime can be understood as an emergent process from the deep neural network of quantum states. The boundary theory retains information on the gauge degrees of freedom of the bulk theory. We find that the emergence of a holographic gravitational theory is connected to a deep learning architecture encoded in the quantum field theory. Deep learning works here because it is based on a computational mechanism similar to the mathematical formulation of the Holographic Principle \cite{witten, susskind}. 

Deep neural networks contain multiple layers of hidden neurons in comparison with shallow neural networks and can better model physical systems. Same as in RG, at a critical point, for a better multiscale approximation of the physical system, the number of hidden neurons becomes higher.

\section{Deep learning mapping model of Renormalization Group}

The first step of the Renormalization Group decreasing the resolution by changing
the minimum length scale from a smaller scale to a larger one.
This is done by integrating out fluctuations of the fields
that occur on length scales finer than the renormalization scale.
The final result is a renormalisation of the Hamiltonian which leads
to an effective Hamiltonian expressed in terms of a coarse-grained field.
As the Hamiltonian $H$ depends on fields and couplings, the entire system is described by the partition function, a scale dependent \cite{Saremi} effective action functional encoded in the functional integral over the Hamiltonian. 

In a nutshell, Restricted Boltzmann machines (RBM) are algorithms used for dimensionality reduction, classification, regression, collaborative filtering, feature learning and topic modelling. RBMs are one of the fundamental algorithms of deep learning with applications in dimensional reduction, feature extraction or recommender systems using probability distributions. Deep learning algorithms may be extremely helpful to quantum many-body physics through the connection between RBMs and tensor networks. 

Restricted Boltzmann machines are a probabilistic model consisting of two groups of variables: the visible variables and the hidden variables. The main assumption in is that the hidden units are conditionally independent given the visible units. Here, the Restricted Boltzmann machine (RBM) is composed by layers of hidden neurons and visible neurons represented by spins. If the RBM is properly trained, the hidden units learn to extract useful features from training data.

The RBM distribution is defined through a function that assigns a number to each possible state of the visible and hidden variables. The RBM contains an ensemble of $N$ spins \{$v_i$\}. The index $i$ with ($i=1,2 \cdots N$) gives the position of spins $v_i$ in the lattice. The data is given by the probability distribution $P(\{v_i\})$ of $\{v_i\}$ spins with $i=1 \ldots N$. The probability distribution is also called a Gibbs distribution or partition function. The visible neurons are represented by the physical spin variables $\{v_i\}$. The hidden neurons of the neural network contain further new spin variables $\{h_j\}$. The Renormalization Group and the graph an RBM posses very similar architecture, which makes deep learning suitable for modelling such processes. This connection becomes a bridge between deep learning and quantum many-body physics \cite{JG, GD}. 
The RG transformation generates a deep learning like architecture as the spins on each layer are hidden spins during the RG transformation on the previous layer.
The hidden spins on the next layer are given by the new Hamiltonian using couplings between the neighbour spins. 
The deep learning architecture is described by spins coupled between consecutive layers. 
Correlations between the spins $\{ v_i\}$ generate correlations between the coarse-grained spins $\{ h_j\}$.
We find that RBMs self organize to produce a coarse-graining procedure extremely similar of Kadanoff block renormalization. 
The iterative coarse-graining architecture may be responsible for the great success of deep learning in general. Each new top layer in the RBM learns exponentially higher level features from the input data.  The initial layers successively detect and extract low-level irrelevant features and feed them to the higher layers, while simultaneously learning relevant ones, eventually emerging into an efficient, optimized representation of the initial data.

The new hidden spin variables $\{h_j\}$ ($j=1\ldots M$) are coupled to the visible spins and the interactions between them are described by the energy function
\be
{\mathbf E}(\{v_i\}, \{h_j\}) = \sum_i b_j h_j + \sum_{ij} v_i w_{ij} h_j + \sum_{i} c_i v_i,
\ee
with $\lambda=\{b_j, w_{ij}, c_i\}$ are variational parameters of the model. 

The probability of observing the full system composed of both hidden and visible spins is
\be
p_{\lambda}(\{v_i\}, \{h_j\})= {e^{-{\mathbf E}(\{v_i\}, \{h_j\})} \over \mathcal{Z}}. 
\label{jointprob}
\ee

The visible spins are described by the distribution
\be
p_{\lambda} (\{v_i\})= \sum_{\{h_j\}} p_{\lambda}(\{v_i\}, \{h_j\}) = {\mathrm Tr}_{h_j} p_{\lambda}(\{v_i\}, \{h_j\})
\ee
and the hidden spins by:
\be
p_{\lambda} (\{h_j\})= \sum_{\{v_j\}} p_{\lambda}(\{v_i\}, \{h_j\})={\mathrm Tr}_{v_i} p_{\lambda}(\{v_i\}, \{h_j\}).
\label{marginalhidden}
\ee

At thermal equilibrium, the probability of a spin configuration is calculated using the Boltzmann distribution
\begin{eqnarray}
\nonumber
P(\{v_i\})=\frac{e^{-\mathbf H(\{v_i\})}}{Z}=\\
=\frac{e^{ \sum_i K_i v_i +\sum_{ij} K_{ij} v_i v_j + \sum_{ijk} K_{ijk} v_i v_j v_k +\cdots}}{Z},
\end{eqnarray}
where $H(\{v_i\})$ is the Hamiltonian, and $Z$ is the partition function
\begin{equation}
Z=\mathrm Tr_{v_i}e^{-\mathbf H(\{v_i\})}\equiv \sum_{v_1,\cdots v_N=\pm 1}e^{-\mathbf H(\{v_i\})}.
\end{equation}
where $\mathbf K=\{K_s  \}$ is the coupling constant. 
The Hamiltonian depends on the set of couplings ${\mathbf K} = \{ K_s \}$, describing the set of all possible Hamiltonians.
The spins $\{h_j\}$ are coarse-grained.
The coupling constant is renormalized as $ \tilde{\mathbf K}=\{\tilde K_s  \}$. 
The auxiliary spins $\{h_j\}$ are coupled to physical spins $\{ v_i \}$ and the visible spins have to be averaged out in order to give a coarse-grained picture of the physical spin system. 

The free energy of the spin system is given by
 \be
 F^{v}= -\log{Z} = -\log{\left({\mathrm Tr}_{v_i} e^{-{\mathbf  H}(\{ v_i\} )}\right)}.
 \ee
while the free energy of the coarse grained spin system is
\be
 F_\lambda^h=  -\log{\left({\mathrm Tr}_{h_i} e^{-{{\mathbf  H}^{RG}}_\lambda(\{ h_i\} )}\right)}.
 \ee

The Renormalization group is generated by repeating the computational step of solving the effective Hamiltonian in order to obtain multiple effective spins. This step becomes depth in deep learning. The coarse-grained Hamiltonian is
\begin{eqnarray}
\nonumber
\mathbf H^{RG}(\{h_j\})=\\
-\sum_i \tilde K_i h_i -\sum_{ij} \tilde K_{ij} h_i h_j - \sum_{ijk} \tilde K_{ijk} h_i h_j h_k +\cdots,
\end{eqnarray}
Each step of the Renormalization group represents a layer in the deep neural network.
The adjustable function $\mathbf T_\lambda (\{v_i\},\{h_j\})$ has the parameters $\{\lambda \}$.

The Hamiltonian $  \mathbf{H}^{RG}_\lambda[\{h_j\}]$ of the coarse-grained RG degrees of freedom is also the Hamiltonian for the hidden spins in a deep neural network. 
\be
\mathbf{H}^{RG}_\lambda[\{h_j\}]=\mathbf{H}^{RBM}_\lambda[\{h_j\}].
\ee

The RG transformation provides the approximate solution, dependent on the function ${\mathbf T}_\lambda(\{v_i \},\{h_j\})$
\begin{equation}
e^{-\mathbf H^{RG}(\{h_j\})}\equiv \mathrm Tr_{v_i}e^{\mathbf T_\lambda (\{v_i\},\{h_j\})-\mathbf H(\{v_i\})},
\end{equation}
The parameter $\lambda$ minimizes the free energy difference $\Delta F = F_\lambda^h-F^v$ between physical and coarse grained spins and the long-distance physical observables remain invariant.
\be
\Delta F=0 \iff  {\mathrm Tr}_{h_j} e^{{\mathbf T}_\lambda(\{v_i\}, \{h_j\})}=1
\ee
The parameters $\lambda=\{b_j, w_{ij}, c_i\}$ are adjustable. The correlations between the hidden spins are given by $\{ \tilde{K} \}$. 
The probability of a configuration of hidden neurons is
\begin{eqnarray}
\nonumber
p_\lambda(\{h_j\})= \sum_{\{v_i\}}\frac{e^{-\mathbf E(\{v_i\},\{h_j\})}}{\mathcal Z}=\\
\sum_{\{v_i\}}\frac{e^{-\sum_j b_j h_j - \sum_{ij} v_i w_{ij} h_j -\sum_i c_i v_i}}{\mathcal Z},
\end{eqnarray}
The couplings are mapped as $\{ K\} \to \{ \tilde{K} \}$.
The Hamiltonian for hidden neurons in the RBM algorithm is
\begin{equation}
p_\lambda(\{h_j\})= \frac{e^{-\mathbf H^{RBM}_{\lambda}(\{h_j\})}}{\mathcal Z}.
\end{equation}
The conditional probability distribution $p_\lambda(\{ h_j\} | \{v_i\})$ can map the layer of visible spins to the hidden spins layer. Once trained, the network uses the hidden layer as a feedback response to a visible new layer as input data for learning the next hidden spins layer. 

The Renormalization group and the deep neural network are directly mapped
\begin{align}
\mathbf T_\lambda (\{v_i\},\{h_j\})&=-\mathbf E(\{v_i\},\{h_j\})+\mathbf H(\{v_i\}), \\
\mathbf H^{RG}_{\lambda}(\{h_j\})&=\mathbf H^{RBM}_{\lambda}(\{h_j\}).
\end{align}
The ${\mathbf T}_\lambda(\{v_i \},\{h_j\})$ is dependent on the set of parameters $\{\lambda\}$ and describes the statistical correlations interactions between the physical and coarse-grained degrees of freedom. 
The function ${\mathbf T}_\lambda(\{v_i\}, \{h_j\})$ is an approximation of the conditional probability of hidden spins based on visible spins.
In this way, we can see that the Renormalization Group has a full natural interpretation in terms of probability theory. 

The couplings between hidden and visible spins are given by ${\mathbf T}_\lambda(\{v_i\}, \{h_j\})$.
The energy function ${\mathbf E}(\{v_i\}, \{h_j\})$ plays the role of the RG coupling between hidden and visible spins. 
For each RG transformation we have
\be
{\mathrm Tr}_{h_j} e^{{\mathbf T}_\lambda(\{v_i\}, \{h_j\})}=1
\label{exactRG}
\ee

We use a short-ranged RBM because only the nearby nodes are connected, at the same time exhibiting area-law entanglement encoded in the graph of the neural network. For a number $\tilde N$ of nodes given by physical spins in a sub-region $A$ of the network, the number of steps  is $\tau= \log_2 \tilde N$. Every layer has a link that must be removed, in order to separate $A$ from the rest of the network. As $A$ is coarse-grained to one node, the minimum number of links to be removed is $n \propto\tau$. 

If we treat the neural network as a tensor network, the entanglement entropy $S_A$ of $A$ is proportional to the number of removed links. The scaling takes the form of an area law, stating that the entanglement entropy of a sub-region $A$ grows with the area  of the boundary of the sub-region rather than its volume. 

Entanglement entropy has been extensively studied in low-dimensional quantum many-body systems to help understanding the nature of quantum criticality. Here, a deep learning view on entanglement entropy can also help with studying critical systems. The line that separates $A$ from the rest of the network can be seen as a Ryu-Takayanagi surface in AdS/CFT. 

We therefore have
\be
S_A \propto c n \propto c\tau= c \log_2 \tilde N,
\ee
that has the same form with the logarithm law for one dimensional systems. Here $c$ is a proportional constant related to the logarithm of the dimensions of the Hilbert space of every link. 

The measure of entanglement $S(A)$ is the mutual information $I(A, B)$ \cite{koch} between two disjoint adjacent subsystems $A$ and $B$.
In the holographic bulk the mutual information has a geometric interpretation where the entanglement entropy $S(A)$ of the region $A$ is proportional to the area of the minimal surface $\gamma_A$. In AdS$_3$, this area is mapped to the geodesic line between two boundary points in the region $A$. 
The entanglement entropy will be given by the single systems
\be
S(A\cup B)=S(A)+S(B)-I(A, B) 
\ee
If $I(A, B)$ is small, the entropy is mainly dominated by the additive part 
$S(A\cup B)\simeq S(A)+S(B)$. 

The mutual information of the two subsystems becomes
\be
I(A,B)=S(A)+S(B)-S(AB)=\frac{1}{4G_N}(|\gamma_{A}|+|\gamma_{B}|-|\gamma_{AB}|)
%$\gamma_{A}$, $\gamma_{B}$ and $\gamma_{AB}$
\ee
and captures the total amount of correlations between the two systems.

In holographic geometric interpretation, the formula describes the three sides of a triangle in the bulk and shows how much is the sum of the two sides greater than the third side. In a semi-classical approach, any disjoint systems which are strongly correlated contain multi-partite entanglement and therefore the mutual information that the formula predicts is always smaller than the physical one. 
Any holographic state will have less mutual information between adjacent systems.

Landau-Fermi liquids do not have any gravity duals in the classical limit while the holographic states discussed here contain
large negative tripartite information that will change the entropy of the system. In contrast, the ground state of the free fermions must be described by a different geometry, while their large mutual information may be understood by large quantum fluctuations of the geometry of the system. Therefore, the geometry fluctuation can explain the deficit of mutual information for strongly correlated states with a holographic dual.
The mutual information also offers an upper bound for the connected two-point functions between the operators
associated with the systems $A$ and $B$.

But so far, the system did not contain any matter. Let's suppose we introduce in the system a fermion of mass $m$. 
The mass of the fermion provides a bound for the IR region at the scale $z_\text{IR}\sim\ln\xi$. 
The correlation length $\xi$ of the fermion is finite and is given by
\be
\xi^{-1}=\frac{1}{2}\ln\frac{1+|m|}{1-|m|}
\ee
At the same scale, the entanglement between adjacent regions is described by a high energy theory, in UV. The fermion mass will stop the network to grow deeper at a boundary $z_\text{IR}=\log_2\xi$ where the holographic space terminates.
The AdS$_3$ geometry emerges as the mass decreased and moves the boundary toward the CFT$_2$.
As the fermion mass increases, the scale $z_\text{IR}$ fades away from large $z$ in IR towards small $z$ in UV.

The partition function of the boundary CFT field $\{v_i\}$ has the deep neural representation
\begin{eqnarray}
\nonumber
\mathcal Z= \sum_{\{v_i\},\{h_j\}}e^{-\mathbf E(\{v_i\},\{h_j\})}=\\
\sum_{\{v_i\},\{h_j\}}e^{-\sum_j b_j h_j - \sum_{ij} v_i w_{ij} h_j -\sum_i c_i v_i}.
\end{eqnarray}
where $\{h_j\}$ is the bulk field. Also $\sum_{ij} v_i w_{ij} h_j$ can be seen as $\int \mathrm d^d x \phi(x) O(x)$ term in GKP-W, where $\phi(x)$ is bulk field and $O(x)$ is boundary field. 

The task here is to identify new $M<N $ coarse-grained binary spins $\{h_j \}$ after all short distance fluctuations on smaller scale are averaged out. The length scale of the spin lattice system is changed, the new degree of fredom creating more space available for each $h_j$ spin.

In a one dimensional spin system with translational invariance, the nodes are formed by spin variables with the first line as the physical spin $\{v_i\}$, and the rest being coarse-grained spins $\{h_j\}$. The network s defined in coordinates $(\tau,x)$. 
A coarse-grained spin is in the same equal-$x$ line, and spins in every step of RG are in the equal-$\tau$ line.

We consider a discrete $d$-dimensional space described by a $d$-dimensional lattice viewed as a fixed graph $\Gamma$ within $\mathbb R^d$. Every vertex $v$ is associated with a site random variable $X_v$ taking any $n$ possible value. The lattice's state is given by the joint probability distribution for all variables.

At thermal equilibrium the state of the system can be encoded in the Hamiltonian, proportional to the logarithm of the probability distribution. The energy function is a sum of local functions, dependent only neighbouring sites and relative to the to the graph is $\Gamma$. 

Predicting approximate properties of the system's state from the compact Hamiltonian is always a complex task. However, short-range correlations can be derived at the cost of computational resources. By integrating out the short distance degrees of freedom, an effective theory for the long-distance ones can be built.
The long distance physics becomes readable from the effective Hamiltonian corresponding to low energy modes in absence of any fluctuation. 

The Renormalization group makes non-trivial predictions of long range correlations described by an effective Hamiltonian on a coarse-grained array of random variables. The array of new variables predicted is described by a new renormalized Hamiltonian.
In order to obtain the exact solution of the model, the integration is iterated, generating a family of new effective Hamiltonians at all scales. 

The full state of the system is described by universal emergent properties that describe its behaviour at a large scale using coarse-grained local order parameters. The large-scale properties of the system are described by thermodynamic variables that predict the thermodynamic phase the system. The procedure is similar to deep learning of complex patterns in order to predict  a set of new features using neighbouring coarse edge data. In the RG sequence, the relevant features of the physical system at large length scales are extracted by integrating out the short distance degrees of freedom. Relevant features become increasingly important while other features, found as irrelevant, will have a diminishing effect on the final representation of the physical system at large scales. In the same way, we can say that the fluctuations are integrated out starting at microscopic scale and progressively iterate for fluctuations at larger scales. 

The hidden spins generate effective interactions among the visible ones, while there is no direct connection between the visible spins. If the number of hidden spins or connections is increased in the RBM, more complex functions of the visible units can be processed.

In order to understand the entanglement entropy behind the scaling process in the network, we divide the visible units into two different sets $v=\{v_{i\in A}, v_{i\in B}\}$. Given a distribution $\Psi_\mathrm{RBM}$ of the visible variables obtained by integrating out the hidden units, the reduced density matrix becomes
\begin{equation}
\rho  = \sum_{\{v_{i\in B}\}} \Psi^{\ast}_\mathrm{RBM}\left(\{v_{i\in A}^{\prime},v_{i\in B}\}\right)\,\Psi_\mathrm{RBM}\left(\{v_{i\in A},v_{i\in B}\}\right).
\label{eq:rho}
\end{equation}

By integrating out the hidden variables, the RBM distribution of the visible spins takes the form
\begin{eqnarray}
\Psi_\mathrm{RBM}\left(v\right) &= & \sum_{ h }e^{-E(v,h) } \nonumber  \\ & = & \prod_{i,j} e^{a_{i} v_{i}}\left(1 + e^{b_{j}+ \sum_{i}v_{i}W_{ij}}\right),  \label{eq:RBM3_P}
\end{eqnarray}

The information content of the distribution $\Psi_\mathrm{RBM}$ is given by the entanglement entropy of the system, which filters the correlations between any two neighbouring regions or partitions. The von Neumann entanglement entropy is $S = -\mathrm{Tr}(\rho \ln \rho)$. The spins that remain uncorrelated show no entanglement entropy for any bipartition, while the long range correlations will increase the entanglement entropy. 

The maximum entanglement entropy of an RBM can be quantified using the bond dimension of the network. Larger bond dimension means more entanglement. In order to model a highly entangled system, the bond dimension (and therefore the memory and computation time) will grow exponentially with the entropy. The maximal entanglement entropy bound is represented by the upper bond dimension $\ln D$, for witch all eigenvalues of the reduced density matrix become $1/D$.

Instead of layers of visible and hidden units, we can think of regions or partitions where the hidden units (spins) are been traced out while the sites (regions) are put in contact with each other.
The network is cut in a bipartition of visible units with two regions $A$ and $B=B_{1}\cup B_{2}$, where $B_{1}$ describes the visible spins that can be reached from $A$, and $B_{2}$ describes the rest. 
The product state of maximally entangled pairs connecting two neighbouring regions.

\begin{equation}
\Psi_\mathrm{RBM}\left(v\right) =\psi\left(\{v_{i\in A}, v_{i\in B_{1}}\}\right)\phi\left(\{v_{i\in B_{1}}, v_{i\in B_{2}}\}\right). \label{eq:WaveABC}
\end{equation}

If $m$ is the number of spins at the interface region of the bipartition, the bond dimension is
\begin{equation}
D=2^{m}
\label{eq:n2D}
\end{equation}

The bond dimension gives an (upper) bound for the maximal entanglement entropy at the bipartition
\begin{equation}
S_{\mathrm{max}} = m\ln 2.
\label{eq:S}
\end{equation}
The entanglement entropy bound is dependent on the number of connections in the RBM.
If each visible spin has $m$ effective connections to the neighbour visible spins, we need $2^{m}$ parameters to describe the system. 

\section{Conclusion}

Tensor networks have been used for studying the holographic duality because their entropy is bounded by the
area law in agreement with the Ryu-Takayanagi holographic entanglement formula. RBM directly mapped to tensor networks provide a strong framework to study holographic principle as they obey the entanglement area law. We saw that if the bond dimension is large, the entanglement entropy obey the Ryu-Takayanagi entropy formula. If the average over the tensor network is described as a partition function of a classical ferromagnetic Ising model, the Ryu-Takayanagi minimal surfaces of can be interpreted as domain walls. The area law provides only an upper bound for the entanglement entropy, corresponding to a maximum saturation of the network. In the case of deep learning, the entropy is used in the learning phase of a neural network.

As the learning process progresses, more hidden nodes are saturated, increasing the impact of relevant units while
dampening the less important ones. The less relevant units can be then eliminated to reduce the memory requirements and computational cost for training the neural network. Such a deep learning approach can provide a well-suited  framework for exploiting and studying the holographic principle, with potential life-long application in AdS/CFT correspondence. While the main motivation for this paper is to better understand holographic duality, we only focused on the one-to-one mapping between two concepts once thought as completely different. 

There are still many unknown factors that limit an RBM network to efficiently represent a many-body state. It would be interesting to study these necessary and sufficient conditions
for an efficient model of quantum states and their entanglement properties, fully constructed by using neural networks.
Of course, as we chose an RBM model, it might be also possible that other types of neural networks
to be a better choice for mapping physical features of many-body physics.

Properties of deep learning networks can be extrapolated and extended to assist with a new information theoretic scenario of holographic duality for a better understanding of locality and Renormalization Group. Entanglement entropy between a bulk region and the boundary in a given tensor network can be better understood under the new umbrella of deep learning/renormalization group direct mapping, consistent with the expectations of AdS/CFT.

\end{document}